\documentclass[aps,prd,showpacs,showkeys,amsmath,amssymb]{revtex4}
\usepackage{amsfonts}
\usepackage{graphicx}
\usepackage{amsmath}
\usepackage{mathrsfs}
\usepackage{tabls}
\usepackage{slashed}
\usepackage{bm}
\usepackage{array}
\usepackage{caption2}
\usepackage{epstopdf}
\usepackage{booktabs}
\usepackage{multirow}
\usepackage{array}

\allowdisplaybreaks

\begin{document}
\title{Heavy hybrid mesons in the QCD sum rule}
\author{Peng-Zhi Huang}
\email{pzhuang@pku.edu.cn}
\author{Shi-Lin Zhu}
\email{zhusl@pku.edu.cn}
\affiliation{Department of Physics
and State Key Laboratory of Nuclear Physics and Technology\\
Peking University, Beijing 100871, China}

\begin{abstract}

We study the spectra of the
hybrid mesons containing one heavy quark ($q\bar{Q}g$)
within the framework of QCD sum rules in the heavy quark limit.
The derived sum rules are stable with the variation of the Borel parameter
within their corresponding working ranges.
The extracted binding energy for the heavy hybrid doublets $H^h(S^h)$ and $M^h(T^h)$
is almost degenerate.
We also calculate the pionic couplings between these heavy hybrid
and the conventional heavy meson doublets using the light-cone QCD sum rule method.
The extracted coupling constants are rather small as a whole.
With these couplings we make a rough estimate of the partial widths of these pionic decay channels.
\end{abstract}
\maketitle
\pagenumbering{arabic}

\section{Introduction}\label{introduction}

Hadron states which can not be accommodated in the conventional quark
model have attracted much interest over the past few
decades, partly due to the great success of quark model in the
classification of hadrons and the calculation of hadronic parameters.
Theoretically, the existence of these unconventional
hadrons may be allowed by Quantum Chromodynamics (QCD),
the widely accepted fundamental theory of the strong
interaction.

These unconventional hadrons include multi-quark states
($qq\bar{q}\bar{q}$, $qqqq\bar{q}$, $\cdots$), glueballs ($gg$,
$ggg$, $\cdots$), and hybrids ($q\bar{q}g$, $qqqg$, $\cdots$).
Some of them are totally ``exotic'', namely their
$J^{PC}$ quantum numbers are excluded by the conventional quark model.
A straightforward analysis of $J^{PC}$ reveals that
$0^{--}$, $0^{+-}$, $1^{-+}$, $2^{+-}$, $\cdots$ are the so called exotic ones.
Several mesons with exotic $J^{PC}=1^{-+}$,
e.g. $\pi_1(1400)$ \cite{1400}, $\pi_1(1600)$ \cite{1600},
have been reported in recent years.
These states are usually interpreted to be hybrid mesons.
The $1^{-+}$ states have been studied in the framework of QCD sum rules in several works,
including their masses \cite{QSRmass} and decay properties \cite{QSRdecay}.
They were also investigated extensively in other theoretical schemes,
such as Lattice QCD, the flux tube model, and AdS/QCD etc.

If the above mentioned hybrid mesons exist,
there should also be hybrid mesons containing one heavy quark ($q\bar{Q}g$)
and heavy hybrid quarkonium  ($Q\bar{Q}g$),
although the $J^{PC}$ quantum number of the former is not exotic.
The hybrid mesons containing one heavy quark and heavy hybrid quarkonium
have been studied in \cite{Govaerts}.
The masses of the hybrid quarkoniums were calculated in the heavy quark limit \cite{zhu1}.
The masses and the pionic couplings to conventional heavy mesons of the hybrids
containing one heavy quark were studied in Ref.~\cite{zhu2}.
In the present work, we study the hybrid mesons containing one heavy quark
in the framework of heavy quark effective theory (HQET) \cite{hqet},
in which the expansion is performed in terms of $1/m_Q$, where $Q$ is the
heavy quark involved. At the leading order of $1/m_Q$,
the HQET Lagrangian respects the heavy quark flavor-spin symmetry,
therefore heavy hadrons form a series of degenerate doublets.
The two members in a doublet share the same quantum number $j_l$, the
angular momentum of the light components.
The two $j_l={1 \over 2}$ $S$-wave conventional heavy mesons form a doublet $(0^-,1^-)$ denoted as $H$ and
the $j_l={1\over 2}/{3 \over 2}$ $P$-wave doublets
$(0^+,1^+)/(1^+,2^+)$ are denoted as $S/T$. We
denote the $j_l={3\over 2}/{5 \over 2}$ $D$-wave doubtlets
$(1^-,2^-)/(2^-,3^-)$ as $M/N$.
As far as the heavy hybrid containing one heavy quark are concerned,
the two $j_l={1\over 2}$ doublets with $P=+$ and $P=-$ are denoted as $S^h$ and $H^h$, respectively.
Similarly, the two $j_l={3\over 2}$ doublets with $P=+$ and $P=-$ are denoted as $T^h$ and $M^h$, respectively.
We study the binding energy and pionic couplings to conventional heavy mesons of these heavy hybrid doublets.
The heavy quark flavor-spin symmetry manifests itself throughout our calculation and makes a distinction between
our present work and that in Ref.~\cite{zhu2}.

We calculate the binding energy and decay constants of these hybrid doublets
using Shifman-Vainshtein-Zakharov (SVZ) sum rules \cite{svz}.
After performing the operator product expansion (OPE)
of the $T$ product of two interpolating currents,
we obtain sum rules which relate the binding energy and decay constants of corresponding hybrid mesons
to expressions containing vacuum condensates parameterizing the QCD nonperturbative effect.
The nonperturbative method used to calculate the pionic couplings
between heavy hybrid mesons and conventional heavy mesons
is the light-cone QCD sum rules (LCQSR) \cite{light-cone}.
Now the OPE of the $T$ product of two interpolating currents sandwiched
between the vacuum and an hadronic state is performed near the light-cone
rather than at a small distance as in SVZ sum rules,
and the QCD nonperturbative effects are included in the light-cone distribution amplitudes of the pion state.

The paper is organized as follows.
We construct the interpolating currents for the doublets $D^h$ ($D=H/S/T/M$) in Sec. \ref{interpolatingcurrents}.
We derive the sum rules for the binding energy and decay constants for these doublets in Sec. \ref{bindingenergy}.
The sum rules for their pionic couplings to the doublets $H$ and $S$ are derived in Sec. \ref{pioniccouplings}.
The last section is a short summary.
The light cone distribution amplitudes of the pion
which are employed in the present calculation are collected in the appendix.

\section{Interpolating currents}\label{interpolatingcurrents}

We adopt the following interpolating currents for the doublets $H^h$ and $M^h$:
\begin{eqnarray}\label{currentsHM}
&&J^{\dag}_{H^h_0}
=\sqrt{\frac{1}{2}}\bar{h}_vig_s\gamma_5\sigma_t\cdot G q\,,\nonumber\\
&&J^{\dag\alpha}_{H^h_1}
=\sqrt{\frac{1}{2}}\bar{h}_vig_s\gamma^\alpha_t\sigma_t\cdot G q\,,\nonumber\\
&&J^{\dag\alpha}_{M^h_1}
=\bar{h}_vg_s\biggl[3G_t^{\alpha\beta}\gamma_\beta+i\gamma^\alpha_t\sigma_t\cdot G\biggl] q\,,\nonumber\\
&&J^{\dag\alpha_1\alpha_2}_{M^h_2}
=\sqrt{\frac{3}{2}}\bar{h}_vg_s\gamma_5\biggl[G_t^{\alpha_1\beta}\gamma_\beta\gamma_t^{\alpha_2}
+G_t^{\alpha_2\beta}\gamma_\beta\gamma_t^{\alpha_1}
-\frac{2}{3}ig^{\alpha_1\alpha_2}_t\sigma_t\cdot G\biggl] q\,,
\end{eqnarray}
where $G_{\alpha\beta}=G_{\alpha\beta}^n\lambda^n/2$ and $h_v(x)=e^{im_Qv\cdot x}\frac{1+\slashed{v}}{2}Q(x)$
is the heavy quark field with 4-velocity $v$.
The subscript $t$ is used to denote the transverseness of the corresponding Lorentz tensors to $v$,
namely
\begin{eqnarray}
&&\gamma_t^\alpha=\gamma^\alpha-\slashed{v}v^\alpha\,,\nonumber\\
&&g_t^{\alpha\beta}=g^{\alpha\beta}-v^\alpha v^\beta\,,\nonumber\\
&&\sigma_t^{\alpha\beta}=\sigma^{\alpha\beta}-\sigma^{\alpha\mu}v_\mu v^\beta-\sigma^{\mu\beta}v_\mu v^\alpha\,,\nonumber\\
&&G_t^{\alpha\beta}=G^{\alpha\beta}-G^{\alpha\mu}v_\mu v^\beta-G^{\mu\beta}v_\mu v^\alpha\,.
\end{eqnarray}

The overlapping amplitudes between the above currents and the corresponding hybrids are defined as
\begin{eqnarray}\label{oaHhMh}
&&\langle 0|J_{H^h_0}(0)|H^h_0(v)\rangle=f_{H^h_0}\,,\nonumber\\
&&\langle 0|J_{H^h_1}^\alpha(0)|H^h_1(v, \lambda)\rangle=f_{H^h_1}\eta_{H^h_1}^\alpha(v, \lambda)\,,\nonumber\\
&&\langle 0|J_{M^h_1}^\alpha(0)|M^h_1(v, \lambda)\rangle=f_{M^h_1}\eta_{M^h_1}^\alpha(v, \lambda)\,,\nonumber\\
&&\langle 0|J_{M^h_2}^{\alpha_1\alpha_2}(0)|M^h_2(v, \lambda)\rangle=f_{M^h_2}\eta_{M^h_2}^{\alpha_1\alpha_2}(v, \lambda)\,,
\end{eqnarray}
where $\eta(v, \lambda)$ is the polarization tensor of the corresponding heavy hybrid.
Apparently these polarization tensors are traceless, symmetric to their Lorentz index
and transversal to $v$: $\eta_{\alpha\alpha_2\cdots\alpha_n}v^\alpha=0$.
Furthermore, summations on $\lambda$ give the following projection operators:
\begin{eqnarray}\label{poHhMh}
&&\sum_\lambda\eta^\alpha(v, \lambda)\eta^\beta(v, \lambda)=-g_t^{\alpha\beta}\,,\nonumber\\
&&\sum_\lambda\eta^{\alpha_1\alpha_2}(v, \lambda)\eta^{\beta_1\beta_2}(v, \lambda)
=\frac{1}{2}g_t^{\alpha_1\beta_1}g_t^{\alpha_2\beta_2}
+\frac{1}{2}g_t^{\alpha_1\beta_2}g_t^{\alpha_2\beta_1}
-\frac{1}{3}g_t^{\alpha_1\alpha_2}g_t^{\beta_1\beta_2}\,.
\end{eqnarray}


Now it is straightforward to give the interpolating currents for the doublets $S^h$ and $T^h$
by adding $\gamma_5$ to the currents in Eq. (\ref{currentsHM}):
\begin{eqnarray}\label{currentsST}
&&J^{\dag}_{S^h_0}
=\sqrt{\frac{1}{2}}\bar{h}_vig_s\sigma_t\cdot G q\,,\nonumber\\
&&J^{\dag\alpha}_{S^h_1}
=\sqrt{\frac{1}{2}}\bar{h}_vig_s\gamma_5\gamma^\alpha_t\sigma_t\cdot G q\,,\nonumber\\
&&J^{\dag\alpha}_{T^h_1}
=\bar{h}_vg_s\gamma_5\biggl[3G_t^{\alpha\beta}\gamma_\beta+i\gamma^\alpha_t\sigma_t\cdot G\biggl] q\,,\nonumber\\
&&J^{\dag\alpha_1\alpha_2}_{T^h_2}
=\sqrt{\frac{3}{2}}\bar{h}_vg_s\biggl[G_t^{\alpha_1\beta}\gamma_\beta\gamma_t^{\alpha_2}
+G_t^{\alpha_2\beta}\gamma_\beta\gamma_t^{\alpha_1}
-\frac{2}{3}ig^{\alpha_1\alpha_2}_t\sigma_t\cdot G\biggl] q\,.
\end{eqnarray}
The corresponding overlapping amplitudes and projection operators
can be defined similarly to Eq. (\ref{oaHhMh}) and ({\ref{poHhMh}), respectively.

\section{Binding energy}\label{bindingenergy}

To derive the sum rules for the binding energy for the doublets $H^h$ and $M^h$,
we consider the following correlation functions:
\begin{eqnarray}
&&i\int d^4 xe^{ik\cdot x}\langle 0|T\{J_{H^h_0}(x)J^{\dag}_{H^h_0}(0)\}|0\rangle=\Pi_{H^h_0}(\omega)\,,\nonumber\\
&&i\int d^4 xe^{ik\cdot x}\langle 0|T\{J^\alpha_{H^h_1}(x)J^{\dag\beta}_{H^h_1}(0)\}|0\rangle
  =-g_t^{\alpha\beta}\Pi_{H^h_1}(\omega)\,,\nonumber\\
&&i\int d^4 xe^{ik\cdot x}\langle 0|T\{J^\alpha_{M^h_1}(x)J^{\dag\beta}_{M^h_1}(0)\}|0\rangle
  =-g_t^{\alpha\beta}\Pi_{M^h_1}(\omega)\,,\nonumber\\
&&i\int d^4 xe^{ik\cdot x}\langle 0|T\{J^{\alpha_1\alpha_2}_{M^h_2}(x)J^{\dag\beta_1\beta_2}_{M^h_2}(0)\}|0\rangle
  =\left[\frac{1}{2}g_t^{\alpha_1\beta_1}g_t^{\alpha_2\beta_2}+\frac{1}{2}g_t^{\alpha_1\beta_2}g_t^{\alpha_2\beta_1}
   -\frac{1}{3}g_t^{\alpha_1\alpha_2}g_t^{\beta_1\beta_2}\right]\Pi_{M^h_2}(\omega)\,,
\end{eqnarray}
where $\omega=2k\cdot v$.
The correlation functions for doublets $S^h$ and $T^h$ are similar to that of $H^h$ and $M^h$, respectively.

At the quark level ($\omega \ll0$),
it is convenient to calculate the above correlators in coordinate space by OPE at a short distance $x\rightarrow 0$.
The fourier transformation to the momentum space is straightforward
after employing the heavy quark propagator in the infinity quark mass limit $m_Q\rightarrow \infty$:
\begin{eqnarray}\label{heavyquarkpropagator}
\langle0|T\{h_v (x)\bar{h}_v(0)\}|0\rangle=\frac{1+\slashed{v}}{2}\int_0^\infty dt\delta(x-vt)\,.
\end{eqnarray}
The quark propagator used in the OPE of $\Pi(\omega)$ is
\begin{eqnarray}\nonumber
\langle0|T\{q (x)\bar{q}(0)\}|0\rangle
=\frac{i\slashed{x}}{2\pi^2x^4}
+\frac{i}{32\pi^2}\frac{\lambda^n}{2}g_sG^n_{\mu\nu}\frac{1}{x^2}(\sigma^{\mu\nu}\slashed{x}+\slashed{x}\sigma^{\mu\nu})
-\frac{\langle\bar qq\rangle}{12}
+\cdots\,.
\end{eqnarray}

We consider the contributions of the condensates with dimension not greater than seven in our calculation.
The terms involving the 6-dimensional four quark condensate appear as
$\alpha_s^2 \langle \bar{q}q\rangle^2$.
It is of high order in $\alpha_s$ so it can be omitted safely.
The Feynman diagrams corresponding to the quark-level calculation are presented in Fig. \ref{fig:FeynDiag}.
The Fock-Schwinger gauge $x_\mu A^\mu(x)=0$ adopted for the external gauge field,
together with the heavy quark propagator in Eq. (\ref{heavyquarkpropagator}),
leads to the leading order Lagrangian of HQET $\mathcal{L}_0=\bar{h}_viv\cdot Dh_v=\bar{h}_viv\cdot \partial h_v$.
This indicates the decoupling of the heavy quark from the external gauge field in the heavy quark limit and greatly simplifies our calculation.

\begin{figure}[!htb]
\captionstyle{flushleft}
\includegraphics[width=5.5in]{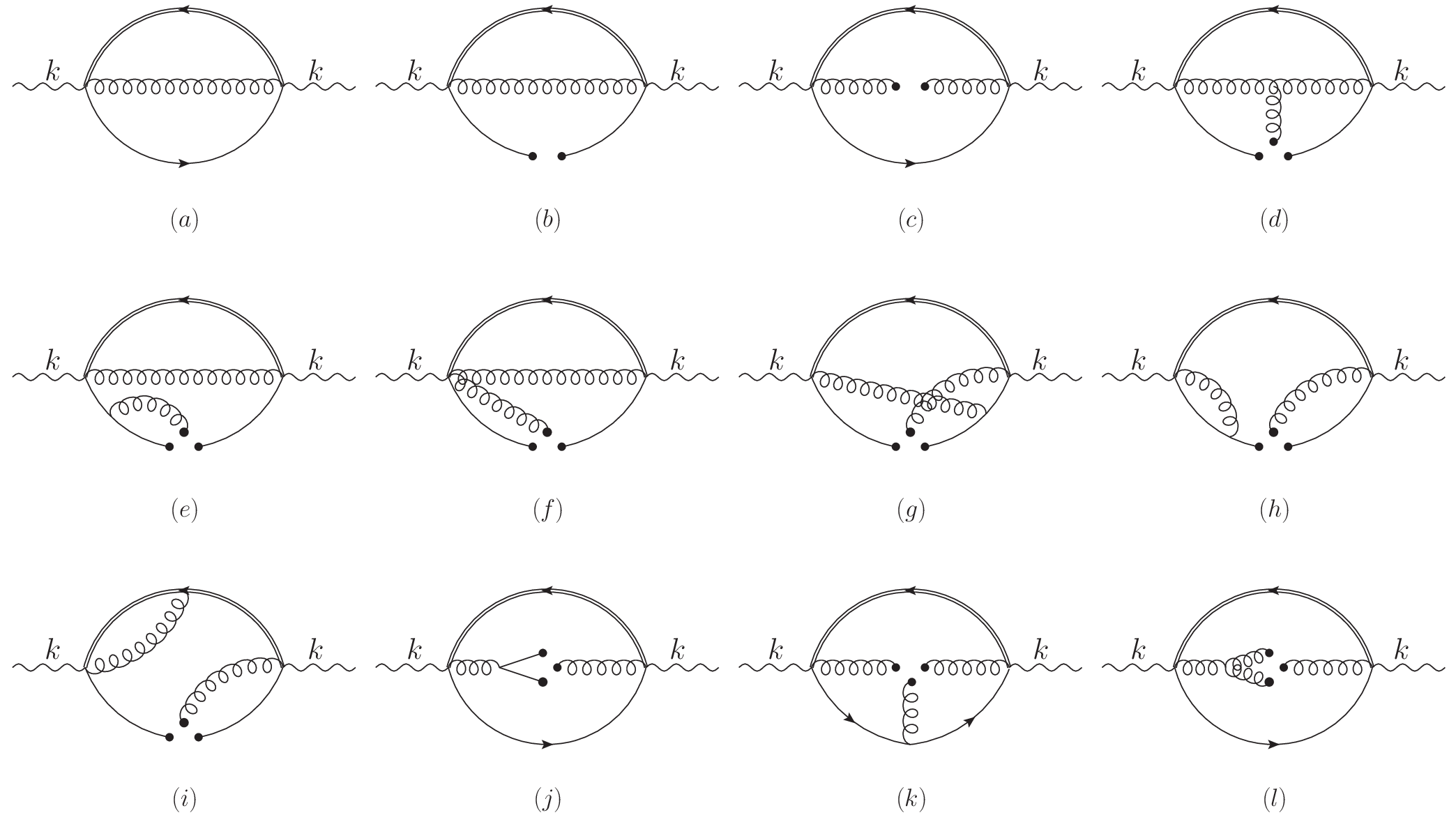}
\caption{The Feynman diagrams for $\Pi(\omega)$. The double solid line denotes the propagator of heavy quark $Q$.}
\label{fig:FeynDiag}
\end{figure}

The OPE results for $\Pi(\omega)$ read
\begin{eqnarray}\label{opeHM}
&&\Pi_{H^h_0}(\omega)
=\int_0^\infty dte^{\frac{it\omega}{2}}
 \left\{-\frac{96\alpha_s}{\pi^3}\frac{1}{t^7}-\frac{16i\alpha_s}{\pi}\frac{\langle\bar{q}q\rangle}{t^4}
 -\frac{1}{4\pi^2}\frac{\langle GG\rangle}{t^3}+\frac{i\alpha_s}{3\pi}\frac{\langle \bar{q}Gq\rangle}{t^2}-\frac{3}{32\pi^2}\frac{\langle GGG\rangle}{t}
 -\frac{i}{24}\langle\bar{q}q\rangle\langle GG\rangle\right\}\,,\nonumber
\\
&&\Pi_{M^h_1}(\omega)
=\int_0^\infty dte^{\frac{it\omega}{2}}
 \left\{-\frac{96\alpha_s}{\pi^3}\frac{1}{t^7}-\frac{16i\alpha_s}{\pi}\frac{\langle\bar{q}q\rangle}{t^4}
 -\frac{1}{4\pi^2}\frac{\langle GG\rangle}{t^3}+\frac{29i\alpha_s}{24\pi}\frac{\langle \bar{q}Gq\rangle}{t^2}-\frac{3}{64\pi^2}\frac{\langle GGG\rangle}{t}
 -\frac{i}{24}\langle\bar{q}q\rangle\langle GG\rangle\right\}\,,\nonumber\\
\end{eqnarray}
and similarly
\begin{eqnarray}\label{opeST}
&&\Pi_{S^h_0}(\omega)
=\int_0^\infty dte^{\frac{it\omega}{2}}
 \left\{-\frac{96\alpha_s}{\pi^3}\frac{1}{t^7}+\frac{16i\alpha_s}{\pi}\frac{\langle\bar{q}q\rangle}{t^4}
 -\frac{1}{4\pi^2}\frac{\langle GG\rangle}{t^3}-\frac{i\alpha_s}{3\pi}\frac{\langle \bar{q}Gq\rangle}{t^2}-\frac{3}{32\pi^2}\frac{\langle GGG\rangle}{t}
 +\frac{i}{24}\langle\bar{q}q\rangle\langle GG\rangle\right\}\,,\nonumber
\\
&&\Pi_{T^h_1}(\omega)
=\int_0^\infty dte^{\frac{it\omega}{2}}
 \left\{-\frac{96\alpha_s}{\pi^3}\frac{1}{t^7}+\frac{16i\alpha_s}{\pi}\frac{\langle\bar{q}q\rangle}{t^4}
 -\frac{1}{4\pi^2}\frac{\langle GG\rangle}{t^3}-\frac{29i\alpha_s}{24\pi}\frac{\langle \bar{q}Gq\rangle}{t^2}-\frac{3}{64\pi^2}\frac{\langle GGG\rangle}{t}
 +\frac{i}{24}\langle\bar{q}q\rangle\langle GG\rangle\right\}\,,\nonumber\\
\end{eqnarray}
where
$\langle GG\rangle=\langle g_s^2G^n_{\alpha\beta}G^n_{\alpha\beta}\rangle$,
$\langle \bar{q}Gq\rangle=\langle g_s\bar{q}\sigma\cdot Gq\rangle$,
$\langle GGG\rangle=\langle g_s^3f^{abc}G^a_{\alpha\beta}G^b_{\beta\gamma}G^c_{\gamma\alpha}\rangle$,
and we used the following formulas in our calculation:
\begin{eqnarray}
&&\langle g_s^2 G^m_{\alpha\beta}G^n_{\gamma\delta}\rangle
=\frac{\delta^{mn}}{96}\left[g_{\alpha\gamma}g_{\beta\delta}-g_{\alpha\delta}g_{\beta\gamma}\right]\langle GG\rangle\,,\nonumber
\\
&&\langle g_s q_i^aG^n_{\mu\nu}\bar{q}_j^b\rangle
=-\frac{1}{192}(\sigma_{\mu\nu})_{ij}(\frac{\lambda^n}{2})_{ab}\langle \bar{q}Gq\rangle\,,
\\
&&\langle g_s^3f^{abc}G^a_{\mu\nu}G^b_{\alpha\beta}G^c_{\rho\sigma}\rangle
=\frac{1}{24}
\left[g_{\mu\sigma}g_{\alpha\nu}g_{\beta\rho}
+g_{\mu\beta}g_{\alpha\rho}g_{\sigma\nu}
+g_{\alpha\sigma}g_{\mu\rho}g_{\nu\beta}
+g_{\rho\nu}g_{\mu\alpha}g_{\beta\sigma}\right.\nonumber
\\
&&\phantom{\langle g_s^3f^{abc}G^a_{\mu\nu}G^b_{\alpha\beta}G^c_{\rho\sigma}\rangle
=\frac{1}{24}}\left.
-g_{\mu\beta}g_{\alpha\sigma}g_{\rho\nu}
-g_{\mu\sigma}g_{\alpha\rho}g_{\nu\beta}
-g_{\alpha\nu}g_{\mu\rho}g_{\beta\sigma}
-g_{\beta\rho}g_{\mu\alpha}g_{\nu\sigma}\right]\langle GGG\rangle\,.
\end{eqnarray}

It turned out that $\Pi_{H^h_1}(\omega)=\Pi_{H^h_0}(\omega)$, $\Pi_{M^h_2}(\omega)=\Pi_{M^h_1}(\omega)$,
$\Pi_{S^h_1}(\omega)=\Pi_{S^h_0}(\omega)$, $\Pi_{T^h_2}(\omega)=\Pi_{T^h_1}(\omega)$,
and we denote them by $\Pi_{H^h}(\omega)$, $\Pi_{M^h}(\omega)$, $\Pi_{S^h}(\omega)$, and $\Pi_{T^h}(\omega)$, respectively.
This implies that the binding energy and the overlapping amplitudes of the two members of a doublet are degenerate,
which is dictated by the heavy quark flavor-spin symmetry.

For $\Pi(\omega)$ we have the following dispersion relation:
\begin{eqnarray}
\Pi(\omega)
=\int_0^\infty\frac{\rho(s)}{s-\omega-i\epsilon}\ ds\,.
\end{eqnarray}
With the phenomenological spectral density $\rho_{\tiny \mbox{PH}}=f^2\delta(s-2\Lambda)+\cdots$,
we can rewrite the above equation to be
\begin{eqnarray}
\frac{f^2}{2\Lambda-\omega}+\cdots
=\int_0^\infty\frac{\rho_{\tiny \mbox{OPE}}(s)}{s-\omega-i\epsilon}\ ds\,,
\end{eqnarray}
where $\Lambda=m_h-m_Q$ is the binding energy of the heavy hybrid $h$ containing a heavy quark $Q$,
$\rho_{\tiny \mbox{OPE}}(s)$ is the spectral density obtained by OPE at the quark level.
By performing the Borel transformation
\begin{eqnarray}
\mathcal{B}_{\omega}^{T}[f(\omega)]
=\lim_{n\rightarrow\infty}\frac{(-\omega)^{n+1}}{n!}
\left.\left(\frac{d}{d\omega}\right)^nf(\omega)\right\vert_{\omega=-nT}\,,
\end{eqnarray}
which is used to suppress the continuum contribution,
we obtain the following sum rule with the continuum contribution subtracted:
\begin{eqnarray}
f^2e^{-2\Lambda/T}
=\int_0^{s_0}\rho_{\tiny \mbox{OPE}}(s)e^{-s/T}\ ds\,.
\end{eqnarray}
It is convenient to obtain the spectral density $\rho_{\tiny \mbox{OPE}}$ by performing a second Borel transformation
$\mathcal{B}^{1/s}_{-1/T}[f(T)]$ to the right hand side of the above equation:
\begin{eqnarray}\label{massSpectralDensity}
&&\rho_{H^h}(s)
=\frac{\alpha_s}{480\pi^3}s^6-\frac{\alpha_s}{3\pi}\langle\bar{q}q\rangle s^3
 +\frac{1}{32\pi^2}\langle GG\rangle s^2-\frac{\alpha_s}{6\pi}\langle\bar{q}Gq\rangle s-\frac{3}{32\pi^2}\langle GGG\rangle
 -\frac{1}{12}\langle\bar{q}q\rangle\langle GG\rangle\delta(s)\,,\nonumber
\\
&&\rho_{M^h}(s)
=\frac{\alpha_s}{480\pi^3}s^6-\frac{\alpha_s}{3\pi}\langle\bar{q}q\rangle s^3
 +\frac{1}{32\pi^2}\langle GG\rangle s^2-\frac{29\alpha_s}{48\pi}\langle\bar{q}Gq\rangle s-\frac{3}{64\pi^2}\langle GGG\rangle
 -\frac{1}{12}\langle\bar{q}q\rangle\langle GG\rangle\delta(s)\,,\nonumber
\\
&&\rho_{S^h}(s)
=\frac{\alpha_s}{480\pi^3}s^6+\frac{\alpha_s}{3\pi}\langle\bar{q}q\rangle s^3
 +\frac{1}{32\pi^2}\langle GG\rangle s^2+\frac{\alpha_s}{6\pi}\langle\bar{q}Gq\rangle s-\frac{3}{32\pi^2}\langle GGG\rangle
 +\frac{1}{12}\langle\bar{q}q\rangle\langle GG\rangle\delta(s)\,,\nonumber
\\
&&\rho_{T^h}(s)
=\frac{\alpha_s}{480\pi^3}s^6+\frac{\alpha_s}{3\pi}\langle\bar{q}q\rangle s^3
 +\frac{1}{32\pi^2}\langle GG\rangle s^2+\frac{29\alpha_s}{48\pi}\langle\bar{q}Gq\rangle s-\frac{3}{64\pi^2}\langle GGG\rangle
 +\frac{1}{12}\langle\bar{q}q\rangle\langle GG\rangle\delta(s)\,.
\end{eqnarray}
The binding energy and overlapping amplitudes can now be expressed as
\begin{eqnarray}
\Lambda=\frac{\int_0^{s_0}s\rho(s)e^{-s/T}ds}{2\int_0^{s_0}\rho(s)e^{-s/T}ds}\,,\ \ \ \ \ \
f^2=e^{2\Lambda/T}\int_0^{s_0}\rho(s)e^{-s/T}ds\,.
\end{eqnarray}

In our numerical analysis,
we use $\alpha_s =4\pi/(11-2/3n_f)\ln(s_0/2/\Lambda_{{\small \mbox{QCD}}})^2$
with $n_f=4$ and $\Lambda_{{\tiny \mbox{QCD}}}=220\ \text{MeV}$.
The mass of up and down quark is ignored.
The quark condensate and gluon condensate adopt the standard values
$\langle\bar qq\rangle=-(0.225 \ \text{GeV})^3$, $\langle GG\rangle=0.038 \ \text{GeV}^4$.
There are several values for the triple gluon condensate
$\langle GGG\rangle=0.045\ \text{GeV}^6$ \cite{svz}, $0.06-0.1\ \text{GeV}^6$ \cite{gluon3cond2},
$0.4\ \text{GeV}^6$ \cite{gluon3cond3}.
The uncertainty caused by this difference is within
6\% for $\Lambda_{H^h}$ and $\Lambda_{S^h}$.
So we will fix $\langle GGG\rangle$ to be $0.045\ \text{GeV}^6$ in the following analysis.
Notice from Eq. (\ref{opeHM}}) and (\ref{opeST}) that
the only difference between the OPE for $H^h$ and $M^h$
or $S^h$ and $T^h$ is the triple gluon condensate.
We can then conclude that the binding energy
of $H^h$ and $M^h$ are almost degenerate,
so is the case of $S^h$ and $T^h$.

From the requirement that the contribution of terms in the OPE
is at least three times larger than that of the next term,
except for $\langle GGG\rangle$,
we get the lower limit of $T$ denoted by $T_{min}$.
If we require that $T_{max}\geq T_{min}+0.4\ \text{GeV}$
and the pole contribution is at least 20\% of the whole sum rule,
the lower limit of continuum threshold $s_0$ is determined.

\begin{figure}
\begin{minipage}[t]{0.5\linewidth}
\centering
\captionstyle{flushleft}
\includegraphics[width=3in]{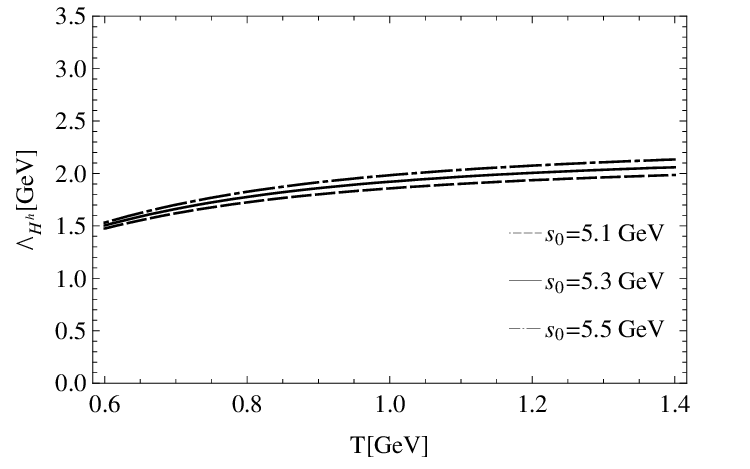}
\setcaptionwidth{3in}
\caption{The sum rule for $\Lambda_{H^h}$ with continuum threshold $s_0=5.1,5.3,5.5\ \text{GeV}$ and the working interval
$0.8<T<1.2\ \text{GeV}$.} \label{fig:BindingEnergyH}
\end{minipage}%
\begin{minipage}[t]{0.5\linewidth}
\centering
\captionstyle{flushleft}
\includegraphics[width=3in]{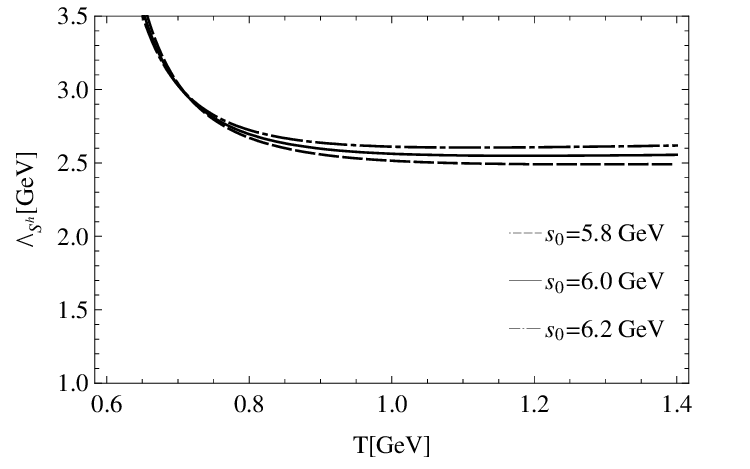}
\setcaptionwidth{3in}
\caption{The sum rule for $\Lambda_{S^h}$ with continuum threshold $s_0=5.8,6.0,6.2\ \text{GeV}$ and the working interval
$0.8<T<1.2\ \text{GeV}$.} \label{fig:BindingEnergyS}
\end{minipage}
\end{figure}

The sum rules for $\Lambda_{H^h}$ and $\Lambda_{S^h}$ are plotted in Fig. \ref{fig:BindingEnergyH}
and Fig. \ref{fig:BindingEnergyS}, respectively.
Finally, we list the extracted the binding energy and the overlapping amplitudes
in Table \ref{massvalue}.
\begin{table}[htb]
\begin{center}
\setlength\extrarowheight{12pt}
\begin{tabular*}{0.5\textwidth}{@{\hspace*{14pt}}@{\extracolsep{\fill}}cccc@{\hspace*{14pt}}}
\hline\ \\[-10mm]
           &  $\Lambda~[\text{GeV}]$  &   $f~[\text{GeV}^{7/2}]$   &   $s_0~[\text{GeV}]$  \\
$H^h/M^h$  &  $2.0$                          &   $1.1$                           &   $5.3$  \\
$S^h/T^h$  &  $2.5$                          &   $1.6$                           &   $6.0$  \\[1mm]
\hline
\end{tabular*}
\end{center}
\caption{The values of $\Lambda$, $f$, and their corresponding $s_0$.}
\label{massvalue}
\end{table}

\section{pionic couplings}\label{pioniccouplings}

To derive the sum rules for the pionic couplings of these heavy hybrids
to conventional heavy mesons,
we need the following interpolating currents
of conventional heavy meson doublets $H$ and $S$:
\begin{eqnarray}
J^\dag_{H_0}&=&\sqrt{\frac{1}{2}}\bar{h}_v\gamma_5 q\,,\nonumber\\
J^{\dag\alpha}_{H_1}&=&\sqrt{\frac{1}{2}}\bar{h}_v\gamma_t^\alpha q\,,\nonumber\\
J^\dag_{S_0}&=&\sqrt{\frac{1}{2}}\bar{h}_v q\,,\nonumber\\
J^{\dag\alpha}_{S_1}&=&\sqrt{\frac{1}{2}}\bar{h}_v\gamma_t^\alpha\gamma_5 q\,.
\end{eqnarray}
The overlapping amplitudes between the above currents and the corresponding heavy mesons are
\begin{eqnarray}\label{oaHM}
&&\langle 0|J_{H_0}(0)|H_0(v)\rangle=f_{H_0}\,,\nonumber\\
&&\langle 0|J_{H_1}^\alpha(0)|H_1(v, \lambda)\rangle=f_{H_1}\epsilon_{H_1}^\alpha(v, \lambda)\,,\nonumber\\
&&\langle 0|J_{S_0}(0)|S_0(v)\rangle=f_{S_0}\,,\nonumber\\
&&\langle 0|J_{S_1}^\alpha(0)|S_1(v, \lambda)\rangle=f_{S_1}\epsilon_{S_1}^\alpha(v, \lambda)\,.
\end{eqnarray}

As an example, we present the derivation of the sum rule for
the coupling constant $g^p_{H^h_1H_1\pi}$,
where $p$ denotes the orbital momentum of the final pion.
$g^p_{H^h_1H_1\pi}$ is defined through the decay amplitude for the channel $H^h_1\rightarrow H_1+\pi$:
\begin{eqnarray}
\mathcal {M}(H^h_1\rightarrow H_1+\pi)
=Ii\varepsilon^{\eta\epsilon^* qv}  g^p_{H^h_1H_1\pi}\,,
\end{eqnarray}
where $q$ is the momentum of the pion,
$\epsilon^*$ is the polarization vector of the final $H_1$ heavy meson,
the isospin factor $I=1, 1/\sqrt{2}$ for the charged and neutral pion, respectively.
$\varepsilon^{\eta\epsilon^*qv}
\equiv\varepsilon^{\mu\nu\rho\sigma}\eta_\mu\epsilon^*_\nu q_\rho v_\sigma$
with $\varepsilon^{\mu\nu\rho\sigma}$ the Levi-Civita tensor.

The correlation function involved in this case is:
\begin{eqnarray}
i\int dx\ e^{-ik\cdot x}\langle\pi(q)|J_{H_1}^\beta(0)J^{\dag\alpha}_{H^h_1}(x)|0\rangle
=Ii\varepsilon^{\alpha\beta\gamma\delta }q_\gamma v_\delta  G^p_{H^h_1H_1\pi}(\omega,\omega')\,,
\end{eqnarray}
where $\omega=2k\cdot v$ and $\omega'=2(k-q)\cdot v$.
When $\omega, \omega' \ll0$,
we work at the quark level and express the above correlation function by the pion light-cone distribution amplitudes:
\begin{eqnarray}
G^p_{H^h_1H_1\pi}(\omega,\omega')
&=&-\frac{1}{2}\int_0^\infty dt\int \mathcal {D}\underline{\alpha}\ e^{it(\frac{\bar{u}}{2}\omega+\frac{u}{2}\omega')}
\frac{f_\pi m_\pi^2}{m_u+m_d}
\biggl\{(m_u+m_d)[2\mathcal{A}_\perp(\underline{\alpha})-\mathcal{V}_\parallel(\underline{\alpha})]
+2\mathcal{T}(\underline{\alpha})q\cdot v\biggr\}\,,\nonumber\\
\end{eqnarray}
where $u\equiv\alpha_2+\alpha_3$ and $\bar{u}\equiv 1-u$.
Furthermore, $G^p_{H^h_1H_1\pi}(\omega,\omega')$ can be related to $g^p_{H^h_1H_1\pi}$
by the dispersion relation
\begin{eqnarray}\label{dispersionrelation}
G^p_{H^h_1H_1\pi}(\omega,\omega')
=\int_0^\infty ds_1\int_0^\infty ds_2 \frac{\rho^p_{H^h_1H_1\pi}(s_1,s_2)}{(s_1-\omega-i\epsilon)(s_2-\omega'-i\epsilon)}
+\int_0^\infty ds_1\frac{\rho^p_1(s_1)}{s_1-\omega-i\epsilon}
+\int_0^\infty ds_2\frac{\rho^p_2(s_2)}{s_2-\omega'-i\epsilon}
+\cdots\,,
\end{eqnarray}
with
\begin{eqnarray}
\rho^p_{H^h_1H_1\pi}(s_1,s_2)
=f_{H^h} f_H g^p_{H^h_1H_1\pi}\delta(s_1-2\Lambda_{H^h})\delta(s_2-2\Lambda_H)+\cdots\,.
\end{eqnarray}

After invoking the double Borel transformation $\mathcal{B}_{\omega}^{T_1}\mathcal{B}_{\omega'}^{T_2}$,
we extract the double dispersion relation part of Eq. (\ref{dispersionrelation}):
\begin{eqnarray}
f_{H^h} f_H g^p_{H^h_1H_1\pi} e^{-2\bar{u}_0\Lambda_{H^h}/T-2u_0\Lambda_H/T}
=f_\pi m_\pi^2
 \left\{\frac{1}{m_u+m_d}\mathcal{T}^{[1]}(u_0)T^2f_1(\frac{\omega_c}{T})
 +\left[\mathcal{V}_\parallel^{[0]}(u_0)-2\mathcal{A}_\perp^{[0]}(u_0)\right]Tf_0(\frac{\omega_c}{T})\right\}\,,
\end{eqnarray}
where
\begin{eqnarray}
u_0=\frac{T_1}{T_1+T_2},\ \ \ \ \ \ T=\frac{T_1T_2}{T_1+T_2}\,.
\end{eqnarray}
The definitions of $\mathcal{F}^{[\alpha_i]}$s are
\begin{eqnarray}
\mathcal{F}^{[0]}(u_0)&\equiv&\int_0^{u_0}\mathcal{F}(\bar{u}_0,\alpha_2,u_0-\alpha_2)\,d\alpha_2\,,\nonumber
\\
\mathcal{F}^{[1]}(u_0)&\equiv&\int_0^{u_0}\frac{\mathcal{F}(\bar{u}_0,\alpha_2,u_0-\alpha_2)}{u_0-\alpha_2}\,d\alpha_2
                         -\int_0^{\bar{u}_0}\frac{\mathcal{F}(\bar{u}_0-\alpha_3,u_0,\alpha_3)}{\alpha_3}\,d\alpha_3\,,\nonumber
\\
\mathcal{F}^{[2]}(u_0)&\equiv&
\left.\frac{\mathcal{F}(\bar{u}_0-\alpha_3,u_0,\alpha_3)}{\alpha_3}\right|_{\alpha_3=0}
+\frac{\mathcal{F}(0,u_0,\bar{u}_0)}{\bar{u}_0}
+\left.\int_0^{u_0}d\alpha_2
 \frac{\partial[\mathcal{F}(1-\alpha_2-\alpha_3,\alpha_2,\alpha_3)/\alpha_3]}{\partial\alpha_3}\right|_{\alpha_3=u_0-\alpha_2}\nonumber\\
&&-\left.\int_0^{\bar{u}_0}d\alpha_3
  \frac{\partial[\mathcal{F}(1-\alpha_2-\alpha_3,\alpha_2,\alpha_3)/\alpha_3]}{\partial\alpha_2}\right|_{\alpha_2=u_0}\,,\nonumber
\\
\mathcal {F}^{[-1]}(u_0)&\equiv&
\int_0^1\int_0^{1-\alpha_2}\mathcal {F}(1-\alpha_2-\alpha_3,\alpha_2,\alpha_3)d\alpha_3d\alpha_2
-\int_0^{u_0}\int_0^{u_0-\alpha_2}\mathcal {F}(1-\alpha_2-\alpha_3,\alpha_2,\alpha_3)d\alpha_3d\alpha_2\,.
\end{eqnarray}
The function $f_n(x)$ which is introduced while subtracting the contribution of continuum is defined as
\begin{eqnarray}
f_n(x)=1-e^{-x}\sum^n_{i=0}\frac{x^i}{i!}\,.
\end{eqnarray}

Here we present the details of the continuum subtraction.
After invoking the first double Borel transformation to the dispersion relation Eq. (\ref{dispersionrelation}) we arrive at
\begin{eqnarray}
\mathcal{B}_{\omega}^{\frac{1}{\sigma_1}}\mathcal{B}_{\omega'}^{\frac{1}{\sigma_2}}G^p_{H^h_1H_1\pi}(\omega,\omega')
=\int_0^\infty ds_1\int_0^\infty ds_2\;e^{-s_1\sigma_1}e^{-s_2\sigma_1}\rho_{H^h_1H_1\pi}(s_1,s_2)\,.
\end{eqnarray}
Now the spectral density $\rho(s_1,s_2)$ can be derived after a second double Borel transformation:
\begin{eqnarray}
\rho_{H^h_1H_1\pi}(s_1,s_2)=\mathcal{B}_{-\sigma_1}^{\frac{1}{s_1}}\mathcal{B}_{-\sigma_2}^{\frac{1}{s_2}}
\mathcal{B}_{\omega}^{\frac{1}{\sigma_1}}\mathcal{B}_{\omega'}^{\frac{1}{\sigma_2}}G^p_{H^h_1H_1\pi}(\omega,\omega')\,.
\end{eqnarray}
According to quark-hadron duality,
we can subtract the contribution of the excited states and the continuum and arrive at
\begin{eqnarray}
f_{H^h} f_H g^p_{H^h_1H_1\pi} e^{-2\bar{u}_0\Lambda_{H^h}/T-2u_0\Lambda_H/T}
=\int_0^{\omega_c} ds_1\int_0^{\omega'_c} ds_2\;e^{-s_1\sigma_1}e^{-s_2\sigma_2}\;
\mathcal{B}_{-\sigma_1}^{\frac{1}{s_1}}\mathcal{B}_{-\sigma_2}^{\frac{1}{s_2}}
\mathcal{B}_{\omega}^{\frac{1}{\sigma_1}}\mathcal{B}_{\omega'}^{\frac{1}{\sigma_2}}G^p_{H^h_1H_1\pi}(\omega,\omega')\,,
\end{eqnarray}
where $\omega_c$ and $\omega'_c$ are the continuum thresholds of the mass rules of the $H^h$ and $H$ doublets, respectively.
The terms of
$\mathcal{B}_{\omega}^{\frac{1}{\sigma_1}}\mathcal{B}_{\omega'}^{\frac{1}{\sigma_2}}G^p_{H^h_1H_1\pi}(\omega,\omega')$
have general form
$cu_0^mT^n=c\sigma_2^m/(\sigma_1+\sigma_2)^{m+n}$. Here we
assume $m,n>0$ to illustrate the procedure of the continuum
subtraction.
\begin{eqnarray}
&&\int_0^{\omega_c} ds_1\int_0^{\omega'_c} ds_2\;e^{-s_1\sigma_1}e^{-s_2\sigma_2}\;
\mathcal{B}_{-\sigma_1}^{\frac{1}{s_1}}\mathcal{B}_{-\sigma_2}^{\frac{1}{s_2}}
\frac{\sigma_2^m}{(\sigma_1+\sigma_2)^{m+n}}\nonumber\\
&&=\int_0^{\omega_c} ds_1\int_0^{\omega'_c} ds_2\;e^{-s_1\sigma_1}e^{-s_2\sigma_2}\;
\frac{1}{\Gamma(m+n)}\left[-\frac{\partial\delta(s_1-s_2)}{\partial s_1}\right]^ms_1^{m+n-1}\nonumber\\
&&=2\int_0^{\omega'_c} ds_+\int_{-s_+}^{s_+} ds_-\;e^{-s_+T}e^{s_-T_-}\;
\frac{(s_+-s_-)^{m+n-1}}{2^m\Gamma(m+n)}\left(\frac{\partial}{\partial s_-}\right)^m\delta(2s_-)\nonumber\\
&&=\frac{T^n}{2^m}\sum^m_{i=0}\frac{m!}{i!(m-i)!}(2u_0-1)^if_{n-1+i}(\frac{\omega'_c}{T})\,,
\end{eqnarray}
where $s_+=(s_1+s_2)/2$, $s_-=(s_2-s_1)/2$, $1/T_-=1/T_1-1/T_2$ and we assume $\omega_c>\omega'_c$\,.
We will work at the symmetry point, namely $T_1=T_2=2T$ and $u_0=1/2$.
This leads to a greatly simplified continuum subtraction:
\begin{eqnarray}
u_0^mT^n&\rightarrow& \frac{T^n}{2^m}\sum^m_{i=0}\frac{m!}{i!(m-i)!}(2u_0-1)^if_{n-1+i}(\frac{\omega'_c}{T})\nonumber\\
&=&u_0^mT^nf_{n-1}(\frac{\omega'_c}{T})\,,
\end{eqnarray}
namely $T^n\rightarrow T^nf_{n-1}(\omega'_c/T)$.
This choice of $u_0$ is based on the consideration that the working interval of the Borel parameter $T$
of the mass sum rules for $H$ and $S$ is about
$0.8<T<1.1\ \text{GeV}$ \cite{HSparameter},
which is very close to that of the the mass sum rules for $D^h~(D=H/S/M/T)$.
This will enable us to subtract the
continuum contribution cleanly, while the asymmetric choice will
lead to the very difficult continuum substraction \cite{asymmetricpoint}.

The binding energy and the overlapping amplitudes extracted in the previous section are
used in our numerical analysis of the sum rules for the above pionic couplings,
together with the following values for $\Lambda_{H/S}$ and $f_{H/S}$ \cite{HSparameter}:
\begin{eqnarray*}
&&\Lambda_H=0.50\ \text{GeV}\,,\ \ \ \ \ \ f_H=0.25\ \text{GeV}^{3/2}\,,\\
&&\Lambda_S=1.15\ \text{GeV}\,,\ \ \ \ \ \ f_S=0.40\ \text{GeV}^{3/2}\,.
\end{eqnarray*}
In this work the $\pi$ decay constant is taken to be $f_\pi=131\ \text{MeV}$.
$\mu_\pi\equiv m_\pi^2/(m_u+m_d)=(1.573\pm 0.174)\ \text{GeV}$ is
given in Ref. \cite{pilcda}.
The parameters appearing in the $\pi$ distribution amplitudes are listed below \cite{pilcda}.
We use the values at the scale $\mu=1\ \text{GeV}$ in our calculation.
\begin{center}
\setlength\extrarowheight{8pt}
\begin{tabular*}{0.8\textwidth}{@{\hspace*{14pt}}@{\extracolsep{\fill}}ccccccccccccc@{\hspace*{14pt}}}
\hline\ \\[-9mm]
  $a_2$\ \  &$\eta_3$\ \  &$\omega_3$\ \   &$\eta_4$\ \  &$\omega_4$\ \  &$h_{00}$\ \  &$v_{00}$\ \  &$a_{10}$\ \  &$v_{10}$\ \  &$h_{01}$\ \  &$h_{10}$\\[-1mm]
  $0.25$\ \ &$0.015$\ \   &$-1.5$\ \       &$10$\ \      &$0.2$\ \       &$-3.33$\ \   &$-3.33$\ \   &$5.14$\ \    &$5.25$\ \    &$3.46$\ \    &$7.03$  \\[1mm]
\hline
\end{tabular*}
\end{center}

\begin{figure}
\begin{minipage}[t]{0.5\linewidth}
\centering
\captionstyle{flushleft}
\includegraphics[width=3in]{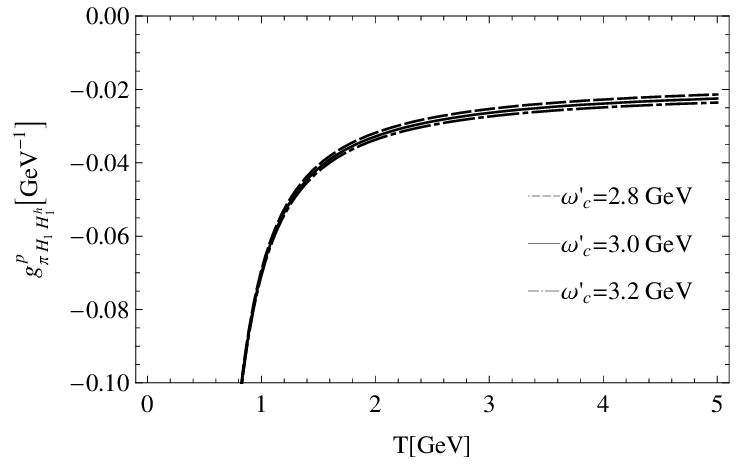}
\setcaptionwidth{3in}
\caption{The sum rule for $g_{H^h_1H_1\pi}^p$ with continuum threshold $\omega'_c=2.8, 3.0, 3.2\ \text{GeV}$.}
\label{fig:ghH1H1p}
\end{minipage}%
\begin{minipage}[t]{0.5\linewidth}
\centering
\captionstyle{flushleft}
\includegraphics[width=3in]{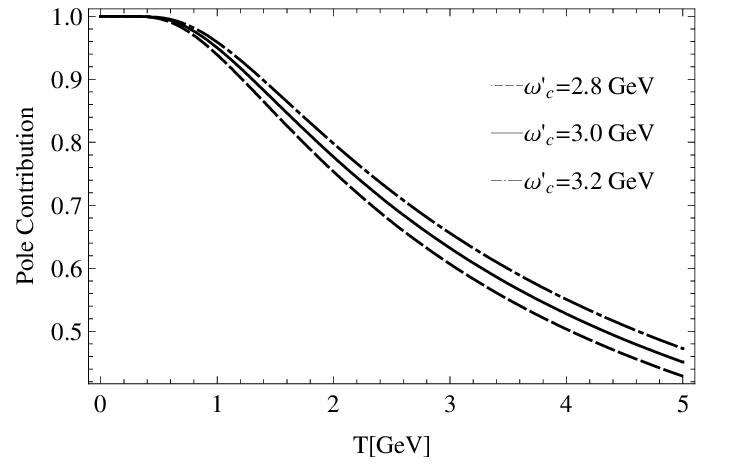}
\setcaptionwidth{3in}
\caption{The determination of the upper limit of $T$ of the sum rule for $g_{H^h_1H_1\pi}^p$,
with the continuum threshold $\omega'_c=2.8, 3.0, 3.2\ \text{GeV}$.}
\label{fig:TmaxH1H1}
\end{minipage}
\end{figure}

From the requirement of the stability of the coupling constant to the variation of the Borel parameter $T$
and the requirement that the pole contribution is larger than 40\%,
we get the working interval of $T_{min}<T<T_{max}$, which is plotted in Fig. \ref{fig:TmaxH1H1}.
The resulting sum rule is plotted with $\omega'_c=2.8, 3.0, 3.2\
\text{GeV}$ in Fig. \ref{fig:ghH1H1p}.

Similarly, we have
\begin{eqnarray}
&&f_{S^h} f_S g^p_{S^h_1S_1\pi} e^{-2\bar{u}_0\Lambda_{S^h}/T-2u_0\Lambda_S/T}
=f_\pi m_\pi^2
 \left\{\frac{1}{m_u+m_d}\mathcal{T}^{[1]}(u_0)T^2f_1(\frac{\omega'_c}{T})
 -\left[\mathcal{V}_\parallel^{[0]}(u_0)-2\mathcal{A}_\perp^{[0]}(u_0)\right]Tf_0(\frac{\omega'_c}{T})\right\}\,,\nonumber
\\
&&f_{M^h} f_H g^p_{M^h_1H_1\pi} e^{-2\bar{u}_0\Lambda_{M^h}/T-2u_0\Lambda_H/T}
=\frac{1}{\sqrt{2}}f_\pi m_\pi^2
 \left\{\frac{1}{2(m_u+m_d)}\mathcal{T}^{[1]}(u_0)T^2f_1(\frac{\omega'_c}{T})
 -\left[\mathcal{A}_\perp^{[0]}(u_0)+\mathcal{V}_\parallel^{[0]}(u_0)\right]Tf_0(\frac{\omega'_c}{T})\right\}\,,\nonumber
\\
&&f_{T^h} f_S g^p_{T^h_1S_1\pi} e^{-2\bar{u}_0\Lambda_{T^h}/T-2u_0\Lambda_S/T}
=\frac{1}{\sqrt{2}}f_\pi m_\pi^2
 \left\{\frac{1}{2(m_u+m_d)}\mathcal{T}^{[1]}(u_0)T^2f_1(\frac{\omega'_c}{T})
 +\left[\mathcal{A}_\perp^{[0]}(u_0)+\mathcal{V}_\parallel^{[0]}(u_0)\right]Tf_0(\frac{\omega'_c}{T})\right\}\,,\nonumber
\\
&&f_{M^h} f_S g^d_{M^h_1S_1\pi} e^{-2\bar{u}_0\Lambda_{M^h}/T-2u_0\Lambda_S/T}
=-\frac{3}{\sqrt{2}}f_\pi m_\pi^2
 \left\{\frac{1}{m_u+m_d}\mathcal{T}^{[0]}(u_0)Tf_0(\frac{\omega'_c}{T})
 -2\left[\mathcal{V}_\perp^{[-1]}(u_0)+\mathcal{V}_\parallel^{[-1]}(u_0)\right]\right\}\,,\nonumber
\\
&&f_{T^h} f_H g^d_{T^h_1H_1\pi} e^{-2\bar{u}_0\Lambda_{T^h}/T-2u_0\Lambda_H/T}
=-\frac{3}{\sqrt{2}}f_\pi m_\pi^2
 \left\{\frac{1}{m_u+m_d}\mathcal{T}^{[0]}(u_0)Tf_0(\frac{\omega'_c}{T})
 +2\left[\mathcal{V}_\perp^{[-1]}(u_0)+\mathcal{V}_\parallel^{[-1]}(u_0)\right]\right\}\,,\nonumber
\\
&&f_{H^h} f_S g^s_{H^h_1S_1\pi} e^{-2\bar{u}_0\Lambda_{H^h}/T-2u_0\Lambda_S/T}=\frac{1}{2}f_\pi m_\pi^2
 \left\{\frac{1}{m_u+m_d}\mathcal{T}^{[2]}(u_0)T^3f_2(\frac{\omega'_c}{T})
 +\left[2\mathcal{V}_\perp^{[1]}(u_0)-\mathcal{V}_\parallel^{[1]}(u_0)\right]T^2f_1(\frac{\omega'_c}{T})\right.\nonumber\\
&&\left.\phantom{f_{H^h} f_S g^s_{H^h_1S_1\pi} e^{-2\bar{u}_0\Lambda_{H^h}/T-2u_0\Lambda_S/T}=}
 +\frac{4m_\pi^2}{m_u+m_d}\mathcal{T}^{[0]}(u_0)Tf_0(\frac{\omega'_c}{T})
 +4m_\pi^2\left[\mathcal{V}_\perp^{[-1]}(u_0)+\mathcal{V}_\parallel^{[-1]}(u_0)\right]\right\}\,,\nonumber
\\
&&f_{S^h} f_H g^s_{S^h_1H_1\pi} e^{-2\bar{u}_0\Lambda_{S^h}/T-2u_0\Lambda_H/T}=\frac{1}{2}f_\pi m_\pi^2
 \left\{\frac{1}{m_u+m_d}\mathcal{T}^{[2]}(u_0)T^3f_2(\frac{\omega'_c}{T})
 -\left[2\mathcal{V}_\perp^{[1]}(u_0)-\mathcal{V}_\parallel^{[1]}(u_0)\right]T^2f_1(\frac{\omega'_c}{T})\right.\nonumber\\
&&\left.\phantom{f_{S^h} f_H g^s_{S^h_1H_1\pi} e^{-2\bar{u}_0\Lambda_{S^h}/T-2u_0\Lambda_H/T}}
 +\frac{4m_\pi^2}{m_u+m_d}\mathcal{T}^{[0]}(u_0)Tf_0(\frac{\omega'_c}{T})
 -4m_\pi^2\left[\mathcal{V}_\perp^{[-1]}(u_0)+\mathcal{V}_\parallel^{[-1]}(u_0)\right]\right\}\,.\nonumber
\\
\end{eqnarray}
We notice that
\begin{eqnarray}
&&g^p_{H^hH\pi}\equiv g^p_{H^h_1H_1\pi}=-g^p_{H^h_0H_1\pi}=-g^p_{H^h_1H_0\pi}\,,\nonumber\\
&&g^s_{H^hS\pi}\equiv g^s_{H^h_1S_1\pi}=-g^s_{H^h_0S_0\pi}\,,\nonumber\\
&&g^s_{S^hH\pi}\equiv g^s_{S^h_1H_1\pi}=g^s_{S^h_0H_0\pi}\,,\nonumber\\
&&g^p_{S^hS\pi}\equiv g^p_{S^h_1S_1\pi}=g^p_{S^h_0S_1\pi}=-g^p_{S^h_1S_0\pi}\,,\nonumber\\
&&g^p_{M^hH\pi}\equiv g^p_{M^h_1H_1\pi}=\frac{1}{2}g^p_{M^h_1H_0\pi}=-\sqrt{\frac{2}{3}}g^p_{M^h_2H_1\pi}\,,\nonumber\\
&&g^d_{M^hS\pi}\equiv g^d_{M^h_1S_1\pi}=\sqrt{\frac{3}{2}}g^d_{M^h_2S_0\pi}=-\sqrt{6}g^d_{M^h_2S_1\pi}\,,\nonumber\\
&&g^d_{T^hH\pi}\equiv g^d_{T^h_1H_1\pi}=-\sqrt{\frac{3}{2}}g^d_{T^h_2H_0\pi}=\sqrt{6}g^d_{T^h_2H_1\pi}\,,\nonumber\\
&&g^p_{T^hS\pi}\equiv g^p_{T^h_1S_1\pi}=\frac{1}{2}g^p_{T^h_1S_0\pi}=\frac{1}{\sqrt{6}}g^p_{T^h_2S_1\pi}\,.
\end{eqnarray}
These relations are consistent with the heavy quark flavor-spin symmetry.

\begin{figure}
\begin{minipage}[t]{0.5\linewidth}
\centering
\captionstyle{flushleft}
\includegraphics[width=3in]{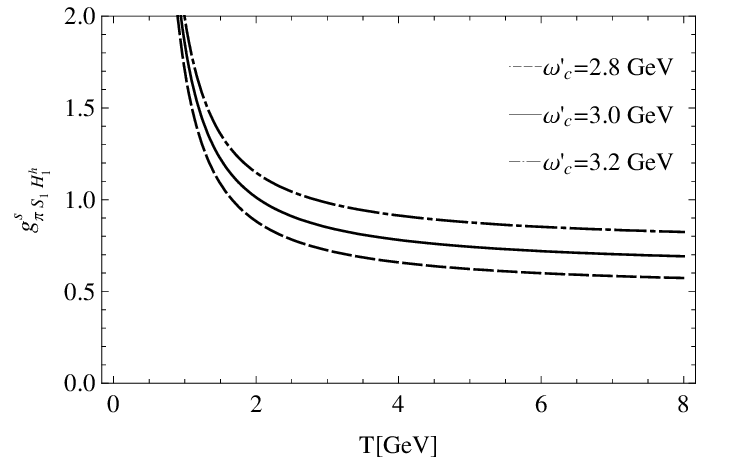}
\setcaptionwidth{3in}
\caption{The sum rule for $g_{H^h_1S_1\pi}^s$ with continuum threshold $\omega'_c=2.8, 3.0, 3.2\ \text{GeV}$.}
\label{fig:ghH1S1s}
\end{minipage}%
\begin{minipage}[t]{0.5\linewidth}
\centering
\captionstyle{flushleft}
\includegraphics[width=3in]{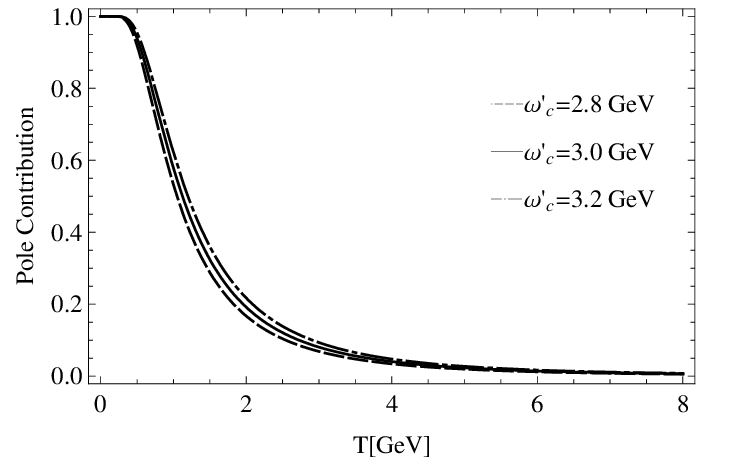}
\setcaptionwidth{3in}
\caption{The upper limit of $T$ of the sum rule for $g_{H^h_1S_1\pi}^s$ is not larger than its lower limit,
indicating the nonexistence of a stable working interval of $T$
($\omega'_c=2.8, 3.0, 3.2\ \text{GeV}$).}
\label{fig:TmaxH1S1}
\end{minipage}
\end{figure}

From Fig. \ref{fig:ghH1S1s} and Fig. \ref{fig:TmaxH1S1}
we can see that there is almost no stable working interval of the Borel parameter $T$,
namely the sum rule for $g_{H^hS\pi}^s$ is unstable,
so is the case of $g_{S^hH\pi}^s$.
Despite this, we make crude estimates for them,
and use these estimates to calculate the partial widths of the corresponding decay channels.

\begin{table}[htb]
\begin{center}
\setlength\extrarowheight{12pt}
\begin{tabular*}{0.8\textwidth}{@{\hspace*{14pt}}@{\extracolsep{\fill}}cccccccc@{\hspace*{14pt}}}
\hline\ \\[-10mm]
  $g_{H^h_1H_1\pi}^p$  &   $g_{S^h_1S_1\pi}^p$   &   $g_{H^h_1S_1\pi}^s$   &   $g_{S^h_1H_1\pi}^s$
& $g_{M^h_1H_1\pi}^p$  &   $g_{T^h_1S_1\pi}^p$   &   $g_{M^h_1S_1\pi}^d$   &   $g_{T^h_1H_1\pi}^d$   \\[-1mm]
  $0.03$                    &   $0.01$                     &   $0.7^*$                    &   $0.7^*$
& $0.2$                     &   $0.1$                      &   $0.08$                     &   $0.1$       \\[1.2mm]
\hline
\end{tabular*}
\end{center}
\caption{The absolute values of the coupling constants.
The units of the $P$- and $D$-wave coupling constants are $\text{GeV}^{-1}$
and $\text{GeV}^{-2}$, respectively.
Here the superscript * indicates the instability of the sum rule for that coupling constant.}
\label{tablecc}
\end{table}

\begin{table}[htb]
\begin{center}
\setlength\extrarowheight{12pt}
\begin{tabular*}{0.9\textwidth}{@{\hspace*{14pt}}@{\extracolsep{\fill}}c@{\hspace*{5pt}}|cccccccc@{\hspace*{14pt}}}
\hline
&  $\ \ H^h_0\ \ $              &  $\ \ H^h_1\ \ $              &  $\ \ S^h_0\ \ $              &  $\ \ S^h_1\ \  $
&  $\ \ M^h_1\ \ $              &  $\ \ M^h_2\ \ $              &  $\ \ T^h_1\ \ $              &  $\ \ T^h_2\ \  $             \\
$\rightarrow H_0^0+\pi^+$
&                               &$\ll0.1$                       & $14.1/24.4$                   &
& $1.9/3.9$                     &                               &                               & $0.1/0.4$                     \\
&                               & $T_p^\alpha$                  & $T_s$                         &
& $T_p^\alpha$                  &                               &                               & $T_d^{\alpha_1\alpha_2}$      \\
$\rightarrow H_1^0+\pi^+$
&$0.03/0.05$                    &$\ll0.1$                       &                               & $17.0/25.2$
& $0.9/1.9$                     & $0.8/1.5$                     & $0.4/1.1$                     & $0.2/0.6$                     \\
& $T_p^\beta$                   & $T_p^{\alpha\beta}$           &                               & $T_s$
& $T_p^{\alpha\beta}$           & $T_p^{\alpha_1\alpha_2\beta}$ & $T_d^{\alpha\beta}$           & $T_d^{\alpha_1\alpha_2\beta}$ \\
$\rightarrow S_0^0+\pi^+$
& $10.7/13.4$                   &                               &                               &$\ll0.1$
&                               &$\ll0.1$                       & $0.5/0.8$                     &                               \\
& $T_s$                         &                               &                               & $T_p^\alpha$
&                               & $T_d^{\alpha_1\alpha_2}$      & $T_p^\alpha$                  &                               \\
$\rightarrow S_1^0+\pi^+$
&                               & $11.0/13.5$                   &$\ll0.1$                       &$\ll0.1$
&$\ll0.1$                       &$\ll0.1$                       & $0.2/0.4$                     & $0.8/1.2$                     \\
&                               & $T_s$                         & $T_p^\beta$                   & $T_p^{\alpha\beta}$
& $T_d^{\alpha\beta}$           & $T_d^{\alpha_1\alpha_2\beta}$ & $T_p^{\alpha\beta}$           & $T_p^{\alpha_1\alpha_2\beta}$ \\[1.5mm]
\hline
\end{tabular*}
\end{center}
\caption{The partial widths of the decay modes $D^h/B^h\rightarrow D/B+\pi$ in unit of MeV,
together with the tensor structures of these decay channels in the heavy quark limit.
The masses of the $c$ quark and $b$ quark used in the calculation are $1.4\ \text{GeV}$ and $4.8\ \text{GeV}$, respectively.}
\label{tablepw}
\end{table}

The extracted coupling constants and the partial widths obtained with them are collected in Table \ref{tablecc}
and Table \ref{tablepw}, respectively.
These numerical values are rather small as a whole.
The annihilation of the gluon degree of freedom in the decay processes may be responsible for these weak couplings.
The tensor structures of the involved decay channels are also included in Table \ref{tablepw},
where the tensor structures of various partial waves are defined as
\begin{eqnarray}
&&T_s=1\,,\nonumber\\
&&T_p^{\alpha}=q_t^\alpha\,,\nonumber\\
&&T_p^{\alpha\beta}=i\varepsilon^{\alpha\beta q v}\,,\nonumber\\
&&T_p^{\alpha_1\alpha_2\beta}=\frac{1}{2}g_t^{\alpha_1\beta}q_t^{\alpha_2}
                             +\frac{1}{2}g_t^{\alpha_2\beta}q_t^{\alpha_1}
                             -\frac{1}{3}g_t^{\alpha_1\alpha_2}q_t^{\beta}\,,\nonumber\\
&&T_d^{\alpha_1\alpha_2}=q_t^{\alpha_1}q_t^{\alpha_2}-\frac{1}{3}g_t^{\alpha_1\alpha_2}q_t^2\,,\nonumber\\
&&T_d^{\alpha_1\alpha_2\beta}=i\varepsilon^{\beta\alpha_1 q v}q_t^{\alpha_2}+i\varepsilon^{\beta\alpha_2 q v}q_t^{\alpha_1}\,.
\end{eqnarray}

\section{Conclusion}\label{summary}

We constructed the appropriate interpolating currents
for the hybrid mesons containing one heavy quark ($q\bar{Q}g$).
Then we calculated the binding energy and the pionic couplings
at the leading order of HQET within the framework of LCQSR.
The mass sum rules and most of the sum rules for the pionic couplings are stable
with the variations of the Borel parameter and the continuum threshold.
For the sum rules for $g_{H^hS\pi}^s$ and $g_{S^hH\pi}^s$,
we can not find a stable working interval of $T$.
We found that the binding energy of the heavy hybrid $H^h$ and $M^h$
are degenerate, so is the case of $S^h$ and $T^h$.
As far as the pionic couplings are concerned,
the extracted couplings are rather small as a whole.

Some possible sources of the errors in our calculation
include the inherent inaccuracy of SVZ sum rules and LCQSR:
the omission of the higher dimensional condensates
in the mass sum rules, and the higher twist terms in the OPE near the light-cone,
the variation of the binding energy and the coupling constant
with the continuum threshold $\omega_c$ and the
Borel parameter $T$ in the working interval, the omission of the
higher conformal partial waves in the light-cone distribution
amplitudes of pion, and the uncertainty in the parameters that appear in these light-cone distribution
amplitudes. The uncertainty in $f$'s and $\bar{\Lambda}$'s is
another source of errors to the light-cone sum rules.
Finally, the $1/m_Q$ correction may turn out to be quite large concerning the charm quark
while such a correction is under control in the case of the bottom quark.

The weak pionic couplings between the heavy hybrid mesons ($q\bar{Q}g$) and the conventional
$q\bar{Q}$ systems render narrow partial widths of the corresponding decay channels.
We have made a rough estimate of these partial widths.
These heavy hybrid mesons are found to be quite narrow
with a width around several tens MeV in the heavy quark limit.
We hope that this estimate,
together with the calculation on the binding energy of the heavy hybrid doublets,
may be helpful to the future experimental search of these unconventional heavy mesons.

\section*{Acknowledgments}

This project is supported by the National Natural Science Foundation of
China under Grants No. 11075004, No. 11021092, No. 11105007, and the Ministry of
Science and Technology of China (No. 2009CB825200).


\appendix

\section{The light-cone distribution amplitudes of the pion}\label{appendixLCDA}

The 2-particle distribution amplitudes of the $\pi$ meson are defined as \cite{pilcda}
\begin{eqnarray}
\langle 0 | \bar u(z) \gamma_\mu\gamma_5 d(-z) |
  \pi^-(P)\rangle
& = & i f_\pi p_\mu \int_0^1 du\, e^{i\xi pz} \, \phi_\pi(u) +
  \frac{i}{2}\, f_\pi m^2\, \frac{1}{pz}\, z_\mu \int_0^1 du \,
  e^{i\xi pz} g_\pi(u)\,,\label{eq:2.8}\nonumber
\\
\langle 0 | \bar u(z) i\gamma_5 d(-z) | \pi(P)\rangle & = &
\frac{f_\pi m_\pi^2}{m_u+m_d}\, \int_0^1 du \, e^{i\xi pz}\,
\phi_{p}(u)\,,
\label{eq:2.11}\nonumber
\\
\langle 0 | \bar u(z) \sigma_{\alpha\beta}\gamma_5 d(-z) |
\pi(P)\rangle & = &-\frac{i}{3}\, \frac{f_\pi
  m_\pi^2}{m_u+m_d}  (p_\alpha z_\beta-
p_\beta z_\alpha) \int_0^1 du \, e^{i\xi pz}\,\phi_{\sigma}(u)\,,
\label{eq:2.12}
\end{eqnarray}
where $\xi\equiv2u-1$, $\phi_\pi$ is the leading twist-2 distribution amplitude, $\phi_{(p,\sigma)}$ are of twist-3.
All the above distribution amplitudes $\phi=\{\phi_\pi,\phi_p,\phi_\sigma,g_\pi\}$
are normalized to unity: $\int_0^1 du\, \phi(u) = 1$.

There is one 3-particle distribution amplitudes of twist-3, defined as \cite{pilcda}
\begin{eqnarray}
\langle 0 | \bar u(z) \sigma_{\mu\nu}\gamma_5
  g_sG_{\alpha\beta}(vz) d(-z)| \pi^-(P)\rangle
& = & i\,\frac{f_\pi m_\pi^2}{m_u+m_d} \left(p_\alpha p_\mu
  g_{\nu\beta}^\perp - p_\alpha p_\nu
  g_{\mu\beta}^\perp - p_\beta p_\mu g_{\nu\alpha}^\perp + p_\beta
  p_\nu g_{\alpha\mu}^\perp \right) {\cal T}(v,pz)\,,\label{eq:3pT3}
\end{eqnarray}
where we used the following notation for the integral
defining the 3-particle distribution amplitude:
\begin{equation}
{\cal T}(v,pz) = \int {\cal D}\underline{\alpha} \, e^{-ipz(\alpha_u
  -\alpha_d + v\alpha_g)} {\cal T}(\alpha_d,\alpha_u,\alpha_g)\,.
\end{equation}
Here $\underline{\alpha}$ is the set of three momentum fractions
$\alpha_d$, $\alpha_u$, and $\alpha_g$. The integration measure is
\begin{equation}
\int {\cal D}\underline{\alpha} = \int_0^1 d\alpha_d d\alpha_u
d\alpha_g \delta(1-\alpha_u-\alpha_d-\alpha_g)\,.
\end{equation}
The 3-particle distribution amplitudes of twist-4 are
\begin{eqnarray}
\langle 0 | \bar u(z)\gamma_\mu\gamma_5
g_sG_{\alpha\beta}(vz)d(-z)|\pi^-(P)\rangle
& = & p_\mu (p_\alpha z_\beta - p_\beta z_\alpha)\, \frac{1}{pz}\, f_\pi
m_\pi^2 {\cal A}_\parallel(v,pz) + (p_\beta g_{\alpha\mu}^\perp -
p_\alpha g_{\beta\mu}^\perp) f_\pi m_\pi^2 {\cal A}_\perp(v,pz)\,,\hspace*{1cm}\nonumber\\
\langle 0 | \bar u(z)\gamma_\mu i
g_s\widetilde{G}_{\alpha\beta}(vz)d(-z)|\pi^-(P)\rangle\
& = & p_\mu (p_\alpha z_\beta - p_\beta z_\alpha)\, \frac{1}{pz}\, f_\pi
m_\pi^2 {\cal V}_\parallel(v,pz) + (p_\beta g_{\alpha\mu}^\perp -
p_\alpha g_{\beta\mu}^\perp) f_\pi m_\pi^2 {\cal V}_\perp(v,pz)\,,\hspace*{1cm}
\end{eqnarray}
where $\widetilde{G}_{\alpha\beta}$ is the dual field
$\widetilde{G}_{\alpha\beta}\equiv \frac{1}{2}\varepsilon_{\alpha\beta\gamma\delta}G^{\gamma\delta}$.

We also use the distribution amplitude given in Ref. \cite{pilcda}:
\begin{eqnarray}
\phi_\pi(u)    &=& 6 u (1-u) \left( 1 + a_2 C_2^{3/2}(\xi)\right)\,,\\
g_\pi(u)       &=& 1+(1 + {18\over7}a_2 + 60\eta_3 + {20\over3}\eta_4)C_2^{1/2}(\xi)
                   + (-{9\over28}a_2 - 6\eta_3\omega_3)C_4^{1/2}(\xi)\,,\\
{\mathbb A}(u) &=& 6u\bar u \left\{ \frac{16}{15} + \frac{24}{35} \,
  a_2 + 20 \eta_3 + \frac{20}{9} \,\eta_4 \right.\nonumber\\
& &\left. + \left( -\frac{1}{15} + \frac{1}{16}\, - \frac{7}{27}\, \eta_3
  \omega_3 - \frac{10}{27}\, \eta_4 \right) C_2^{3/2}(\xi) + \left
  ( -\frac{11}{210} \, a_2 - \frac{4}{135}\, \eta_3\omega_3 \right)
  C_4^{3/2}(\xi) \right\} \nonumber\\
& & {}+ \left(-\frac{18}{5}\, a_2 + 21\eta_4\omega_4 \right) \left\{ 2
  u^3 (10-15 u + 6 u^2)\ln u + 2\bar u^3 (10-15\bar u + 6 \bar u^2)
  \ln\bar u + u \bar u (2 + 13u\bar u)\right\}\,,\\
\mathbb{B}(u) &=& g_\pi(u)-\phi_\pi(u)\,,\\
\phi_p(u)     &=& 1 + \left(30\eta_3 -\frac{5}{2}\, \rho_\pi^2\right)
C_2^{1/2}(\xi) + \left(- 3 \eta_3 \omega_3-\frac{27}{20}\, \rho_\pi^2
  - \frac{81}{10}\, \rho_\pi^2 a_2\right)  C_4^{1/2}(\xi)\,,\\
\phi_\sigma(u)&=& 6u(1-u) \left\{1 + \left(5\eta_3 -\frac{1}{2}\,\eta_3\omega_3 - \frac{7}{20}\,
                  \rho_\pi^2 - \frac{3}{5}\,\rho_\pi^2 a_2 \right) C_2^{3/2}(\xi)\right\}\,,\\
{\cal T}(\underline{\alpha}) &=& 360\eta_3 \alpha_u\alpha_d\alpha_g^2
\left\{1+\omega_3\, \frac{1}{2}\left( 7\alpha_g-3\right)\right\}\,,\\
{\cal V}_\parallel(\underline{\alpha}) & = & 120
\alpha_u\alpha_d\alpha_g ( v_{00} + v_{10} (3\alpha_g-1)),\nonumber\\
{\cal A}_\parallel(\underline{\alpha}) & = & 120
\alpha_u\alpha_d\alpha_g a_{10} (\alpha_d-\alpha_u)\,,\\
{\cal V}_\perp(\underline{\alpha}) & = & -30 \alpha_g^2\left[ h_{00}(1-\alpha_g)
                                         +h_{01}\Big[\alpha_g(1-\alpha_g)-6\alpha_u\alpha_d\Big]
                                         +h_{10}\Big[\alpha_g(1-\alpha_g)-\frac{3}{2}(\alpha_u^2
                                         +\alpha_d^2)\Big]\right]\,,\\
{\cal A}_\perp(\underline{\alpha}) & = & 30 \alpha_g^2(\alpha_u-\alpha_d)\left[ h_{00}+h_{01}\alpha_g
                                            +\frac{1}{2}\,h_{10}(5\alpha_g-3)\right]\,,
\end{eqnarray}
where $C_n^m(\xi)$ are Gegenbauer polynomials.

The definitions and the specific forms  of the $\eta$ light-cone distribution amplitudes adopted in the text are similar to those of the pion.
For more details see Ref. \cite{pilcda}.



\begin{thebibliography}{99}


\bibitem{1400}
D. Alde {\it et~al.}, Phys. Lett. {\bf B205}, 397 (1988);
H. Aoyagi {\it et~al.}, Phys. Lett. {\bf B314}, 246 (1993);
D. R. Thompson {\it et~al.} [E852 Collaboration], Phys. Rev. Lett. {\bf 79}, 1630 (1997);
A. Abele {\it et~al.} [The Crystal Barrel Collaboration], Phys. Lett. {\bf B423}, 175 (1998).

\bibitem{1600}
Yu. P. Gouz {\it et~al.} [VES Collaboration.], AIP Conf. Proc. {\bf 272}, 572 (1993);
Yu A. Khokholov {\it et~al.} [VES Collaboration], Nucl. Phys. {\bf A663}, 596 (2000);
A. Zaitsev {\it et~al.} [VES Collaboration], Nucl. Phys. {\bf A675}, 155c, (2000);
G. S. Adams {\it et~al.} [E852 Collaboration], Phys. Rev. Lett. {\bf 81}, 5760 (1998);
S. U. Chung {\it et~al.} [E852 Collaboration], Phys. Rev. {\bf D65}, 072001 (2002);
E. I. Ivanov {\it et~al.} [E852 Collaboration], Phys. Rev. Lett. {\bf 86}, 3977 (2001);
J. Kuhn {\it et~al.} [E852 Collaboration], Phys. Lett. {\bf B595}, 109 (2004);
M. Lu {\it et~al.} [E852 Collaboration], Phys. Rev. Lett. {\bf 94}, 032002 (2005);
C. A. Baker {\it et~al.}, Phys. Lett. {\bf B563}, 140 (2003);
M. Alekseev {\it et~al.} [The COMPASS Collaboration], Phys. Rev. Lett. {\bf 104}, 241803 (2010);
B. Grube {\it et~al.} [The COMPASS Collaboration], arXiv:1002.1272 [hep-ex] (2010).


\bibitem{QSRmass}
I. I. Balitsky, D. I. D¡¯Yakonov, and A. V. Yung, Phys. Lett. {\bf B112}, 71 (1982);
J. Govaerts {\it et~al.}, Phys. Lett. {\bf B128},  262 (1983);
J. I. Latorre {\it et~al.}, Phys. Lett. {\bf B147}, 169 (1984);
J. Govaerts {\it et~al.}, Nucl. Phys. {\bf B248}, 1 (1984);
I. I. Balitsky, D. I. D¡¯Yakonov, and A. V. Yung, Z. Phys. {\bf C33}, 265 (1986);
J. I. Latorre, S. Narison, and P. Pascual, Z. Phys. C {\bf 34}, 347 (1987);
K. G. Cheyrkin and S. Narison, Phys. Lett. {\bf B485}, 145 (2000);
H. Y. Jin, J. G. Korner, and T. G. Steele, Phys. Rev. {\bf D67}, 014025 (2003);
K. C. Yang, Phys. Rev. {\bf D76}, 094001 (2007);
S. Narison, Phys. Lett. {\bf B675}, 319 (2009).

\bibitem{QSRdecay}
F. De Viron and J. Govaerts, Phys. Rev. Lett. {\bf 53}, 2207 (1984);
J. I. Latorre, P. Pascual, and S. Narison, Z. Phys. {\bf C34}, 347 (1987);
S. L. Zhu, Phys. Rev. {\bf D60}, 097502 (1999);
H. X. Chen, Z. X. Cai, P. Z. Huang, and S. L. Zhu, Phys. Rev. {\bf D83}, 014006 (2011);
P. Z. Huang, H. X. Chen, and S. L. Zhu, Phys. Rev. {\bf D83}, 014021 (2011).



\bibitem{Govaerts}
J.Govaerts {\it et~al.}, Nucl. Phys. {\bf B258}, 215 (1985);
Nucl. Phys. {\bf B262}, 575 (1985); Nucl. Phys. {\bf B284}, 674 (1987).

\bibitem{zhu1}
S. L. Zhu, Phys. Rev. {\bf D60}, 031501 (1999).

\bibitem{zhu2}
S. L. Zhu, Phys. Rev. {\bf D60}, 014008 (1999).

\bibitem{hqet}
B. Grinstein, Nucl. Phys. {\bf B339}, 253 (1990); E. Eichten and
B. Hill, Phys. Lett. {\bf B234}, 511 (1990); A. F. Falk, H.
Georgi, B. Grinstein, and M. B. Wise, Nucl. Phys. {\bf B343}, 1
(1990).


\bibitem{svz}
M.A. Shifman, A.I. Vainshtein, and V.I. Zakharov, Nucl. Phys. {\bf B147}, 385, 448, 519 (1979).

\bibitem{light-cone}
I.I. Balitsky, V.M. Braun, and A.V. Kolesnichenko, Nucl. Phys. {\bf B312}, 509 (1989);
V.M. Braun and I.E. Filyanov, Z. Phys. {\bf C44}, 157 (1989);
V.L. Chernyak and I.R. Zhitnitsky, Nucl. Phys. {\bf B345}, 137 (1990).

\bibitem{gluon3cond2}
A. R. Zhitnitsky, Sov. J. Nucl. Phys. {\bf 41}, 1035 (1985).

\bibitem{gluon3cond3}
I.I.Balitsky, D.I.Dyakanov, and A.V.Yung, Z. Phys. {bf C33}, 265 (1986).

\bibitem{HSparameter}
E. Bagen, P. Ball, V. M. Braun, and H. G. Dosch, Phys. Lett. {\bf B278}, 457 (1992);
M. Neubert, Phys. Rev. {\bf D45}, 2451 (1992);
D. J. Broadhurst and A. G. Grozin, Phys. Lett. {\bf B274}, 421 (1992);
Y. B. Dai and S. L. Zhu, Phys. Rev. {\bf D58}, 074009 (1998).

\bibitem{asymmetricpoint}
I. I. Balitsky, V. M. Braun, and A. V. Kolesnichenko, Nucl. Phys. {\bf B312}, 509 (1989);
V. M. Braun and I. E. Filyanov, Z. Phys. {\bf C44}, 157 (1989);
V. M. Belyaev, V. M. Braun, A. Khodjamirian, and R. Ruckl, Phys. Rev. {\bf D51}, 6177 (1995).



\bibitem{pilcda}
P. Ball, JHEP {\bf 9901}, 010 (1999);
P. Ball, V. M. Braun, and A. Lenz, JHEP {\bf 0605}, 004 (2006).





\end{thebibliography}
\end{document}